\documentclass[aps, pra, showpacs, superscriptaddress, 10pt, twocolumn]{revtex4-1}

\usepackage{amsfonts,amssymb,amsmath,amsthm,longtable,array}
\usepackage{graphics,graphicx,color,enumitem}
\usepackage{epstopdf,epsfig}

\def\identity{\leavevmode\hbox{\small1\kern-3.8pt\normalsize1}}
\def\openone{\leavevmode\hbox{\small1 \normalsize \kern-.64em1}}
\newcommand{\ba}{\begin{align}}
\newcommand{\ea}{\end{align}}
\newcommand{\bpf}{\begin{proof}}
\newcommand{\epf}{\end{proof}}
\newcommand{\ket}[1]{ | #1 \rangle}
\newcommand{\bra}[1]{ \langle #1  |}

\newcommand{\proj}[1]{\ket{#1}\bra{#1}}

\begin{document}

\title{Closing the detection loophole in multipartite Bell experiments with a limited number of efficient detectors}

\author{Kamil Kostrzewa}
\affiliation{Institute of Theoretical Physics and Astrophysics, Faculty of Mathematics, Physics and Informatics, University of Gda\'nsk, 80-308 Gda\'nsk, Poland}

\author{Wies{\l}aw Laskowski}
\affiliation{Institute of Theoretical Physics and Astrophysics, Faculty of Mathematics, Physics and Informatics, University of Gda\'nsk, 80-308 Gda\'nsk, Poland}

\author{Tam\'as V\'ertesi}
\address{Institute for Nuclear Research, Hungarian Academy of Sciences, H-4001 Debrecen, P.O. Box
    51, Hungary}

\begin{abstract}
The problem of closing the detection loophole in Bell tests is investigated in the presence of a limited number of efficient detectors using emblematic multipartite quantum states. To this end, a family of multipartite Bell inequalities is introduced basing on local projective measurements conducted by $N-k$ parties and applying a $k$-party Bell inequality on the remaining parties. Surprisingly, we find that most of the studied pure multipartite states involving e.g. cluster states, the Dicke states, and the Greenberger-Horne-Zeilinger states can violate our inequalities with only the use of two efficient detectors, whereas the remaining detectors may have arbitrary small efficiencies. We believe that our inequalities are useful in Bell experiments and device-independent applications if only a small number of highly efficient detectors are in our disposal or our physical system is asymmetric, e.g. atom-photon.
\end{abstract}

\maketitle

\section{Introduction}

In the famous Einstein-Podolsky-Rosen (EPR) paradox~\cite{EPR}, the authors suggest that quantum mechanics is incomplete. Bell reply to this suggestion~\cite{Bell}, arguing that any local and realistic theory must produce some fundamental bounds for the correlations of results of measurements (given in the form of some inequalities)~\cite{Bellreviews}. In fact, Bell inequalities are one of the most important tools used to detect nonclassical properties of quantum states and certify safety of protocols in a device-independent way in tasks such as quantum cryptography \cite{Crypt}, quantum random number generation \cite{Random} or quantum teleportation \cite{Telep}. Soon after Bell's paper, the first Bell inequality possible to be measured was presented in Ref.~\cite{CHSH} by Clauser-Horne-Shimony-Holt (CHSH). On the
other hand, in Ref.~\cite{CH} a new inequality was proposed by Clauser and Horne (CH), which was even easier to perform in an experiment. Note also the recent paper~\cite{CHCHSH} comparing the power of the CH and CHSH inequalities due to finite statistics relevant in experimental Bell tests.

In Refs.~\cite{FC, A1, A2, A3} pioneering experiments with violation of Bell inequalities were performed, however they were widely discussed, because of some additional assumptions occurring in the tests, which in practice opened some technical loopholes. After then, closing all the technical loopholes or improvement of the efficiency of the experimental devices was one of the most important aims in this research area.

On the theoretical side, asymmetric Bell experiments were under consideration in Refs.~\cite{E0,E1}.  The minimum detection efficiency of $\eta_{crit}=0.43$ can be tolerated for one of the particles, if the other one is always detected.
Closing the detection loophole in Bell experiments using high dimensional systems~\cite{E4} and multipartite systems~\cite{E5,E6} were also addressed. In particular, in the above works the authors find $\eta_{crit} \approx 0.62$ for two-ququart states, $\eta_{crit} \approx 0.38$ for the Greenberger-Horne-Zeilinger (GHZ) states~\cite{GHZ} with a reasonable number of qubits, and $\eta_{crit} \approx 0.50$ for the three-qubit W state~\cite{Wstate} as a minimum detection efficiency needed to violate a Bell inequality.

On the experimental side, in Ref.~\cite{E2} the first Bell test free of detection efficiency loophole for photons was reported. However, the locality loophole was still open in that experiment. Eventually, in 2015 three independent groups reported realisation of a loophole free Bell experiment~\cite{H, G, S}. Shortly after completing these experiments, in Ref.~\cite{R} an experiment closing the locality and the detection loopholes on entangled atoms was also performed.

In this paper we present a method for constructing $N$-particle Bell inequalities based on some known $k$-particle two-setting Bell inequalities, where the latter can be violated with only $k$ pairs of efficient detectors.  The presented approach can be also used in situations when one performs a Bell experiment on two different physical systems. For example, in an atom-photon system, the $k$ atoms can be detected with probability close to 1, whereas the probability of detecting the $N-k$ photons is much smaller.
In section \ref{SecII}, a general approach is presented and we also present some examples for possible violation of this inequality with the use of different quantum states. We discuss the critical detection efficiency threshold in our approach and compare our results with results deriving from known inequalities. In section \ref{time}, we explain how low efficiency of detectors in our scenario influence the duration of the experiment. In section \ref{pers}, we discuss our results in term of persistency of nonlocality. In section \ref{s-mixed} we present the generalization to mixed states. Then,  at the end we give some conclusions.

\section{The method}
\label{SecII}

\begin{figure}
        \centering
    \includegraphics[width=0.46\textwidth]{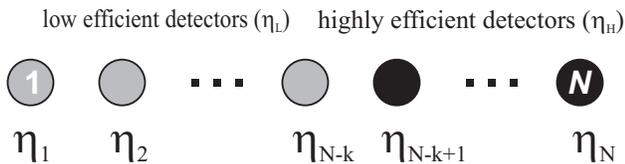}
    \caption{\label{fig-eff} Experimental scenario. The $N-k$ first observers that have inefficient detectors make a projection onto some state. The remaining $k$ observers perform a typical Bell experiment.}
\end{figure}

Let us consider $N$ observers performing measurements on a given state $\rho_N$. In a realistic scenario, the probability that a particle is detected by a detector is equal to $\eta < 1$. The parameter $\eta$ is usually called {\em detection efficiency} and may be different for each of the detectors. In our case (see Fig.~\ref{fig-eff}) the number of efficient detectors is limited. We assume that the first $N-k$ observers have only detectors with low efficiency ($\eta_1 = \ldots =\eta_{N-k} = \eta_L \ll 1$). The detectors of the last $k$ observers, on the other hand, are highly efficient ($\eta_L \ll \eta_{N-k+1} = \ldots = \eta_N = \eta_H$). The task of the first $N-k$ observers is to measure only some arbitrary projectors $\Pi^+_i$ ($i = 1, \dots, N-k$), whereas each of the last $k$ observers measure two alternative observables defined by two orthogonal projectors, $A_i^j = \Pi^+_{i,j} - \Pi^-_{i,j}$ ($i=N-k+1,\dots,N$; $j=1,2$).

In order to exhibit correlations not describable by classical statistics (i.e., by local realistic models) we introduce the following Bell-type inequality
\begin{eqnarray}
p_1  p_2  \cdots p_{N-k} \langle I_{N-k+1,\dots,N} - L \rangle \leq 0,
\label{ineq}
\end{eqnarray}
where $p_i = {\rm Tr} \rho \Pi^+_i$ are probabilities of obtaining a ``+''-result by the $i$th observer and $\langle I_{N-k+1,\dots,N} \rangle \leq L$ is some Bell inequality defined on the last $k$ particles. The inequality (\ref{ineq}) was previously used in \cite{NOISE} to reveal highly noise resistant quantum correlations.

Now we find the critical efficiencies $\eta_i$ to violate the inequality (\ref{ineq}). In our scenario, we consider a typical detection model \cite{gis} that assumes that the probability of detecting a particle in the ``+'' detector is $\eta$ and if the ``+'' detector does not click it means that the ``--'' detector clicks. The corresponding projectors can be written in the following way:
\begin{eqnarray}
\Pi^{+}_{i,j}(\eta) &=& \eta \Pi^{+}_{i,j}, \\
\Pi^{-}_{i,j}(\eta) &=& \openone - \eta \Pi^{+}_{i,j}.
\end{eqnarray}
To calculate the critical efficiencies, we replace all projectors $\Pi$ in (\ref{ineq}) by the corresponding ones $\Pi(\eta)$ and solve the following equation for $\eta_H$:
\begin{eqnarray}
\eta_L^{N-k} p_1 \cdots p_{N-k} \langle I_{N-k+1,\dots,N}^{\eta_H} - L \rangle = 0,
\label{equ}
\end{eqnarray}
where the explicit form of $I_{N-k+1,\dots,N}^{\eta_H}$ depends on the chosen inequality $I_{N-k+1,\dots,N}$.

\subsection{How ineffective can the detectors be?}

Let us first concentrate on the first group of the detectors -- ineffective ones ($\eta_1 = \dots = \eta_{N-k} = \eta_L$). Since we consider only nontrivial solutions of (\ref{equ}), $\eta_L$ should be greater than 0 but can be arbitrarily small. It means that one can use any but not blind detectors. However, if we use detectors with too low efficiency, we have to increase the time of the experiment. This problem of getting a trade-off between the duration of the experiment and available detection efficiencies will be discussed in Sec. \ref{time}.

\subsection{How effective should the detectors be?}

We now consider the second group of the detectors -- effective ones ($\eta_{N-k+1} = \dots = \eta_N = \eta_H$). In order to calculate the critical efficiency $\eta_H$ which violates inequality $I_{N-k+1,\dots,N}\le L$ we have to solve the equation:
\begin{equation}
    \langle I_{N-k+1,\dots,N}^{\eta_H} - L \rangle = 0.
\end{equation}
The quantum value of $I_{N-k+1,\dots,N}^{\eta_H}$ should be calculated for the resulting state $\rho'$ after the projections of the first $N-k$ qubits:
\begin{equation}
\rho' = \frac{\Pi_1^+ \cdots \Pi_{N-k}^+ \rho_N \Pi_1^+ \cdots \Pi_{N-k}^+}{{\rm Tr}(\Pi_1^+ \cdots \Pi_{N-k}^+ \rho_N \Pi_1^+ \cdots \Pi_{N-k}^+)}.
\end{equation}

\subsection{Examples}
\label{SecIII}

We present the performance of our method for prominent families of multipartite states, i.e. the $N$-qubit GHZ state~\cite{GHZ}, the $N$-qubit Dicke state~\cite{Wstate}, the four-qubit cluster state~\cite{cluster} in a situation when $I$ is the well-known CHSH inequality defined on the last two qubits (i.e., $k=2$):
\begin{eqnarray}
I_{N-1,N} &=& \langle A^1_{N-1} A^1_N + A^1_{N-1} A^2_N + A^2_{N-1} A^1_N \nonumber \\ &-& A^2_{N-1}  A^2_N \rangle \leq 2.
\end{eqnarray}
For those states, one can choose the projections $\Pi_i^+$ ($i=1, \dots, N-2$) in such a way that the resulting two-qubit state is just one of the Bell states. Thus the critical detection efficiency $\eta_H^{crit}$ for those states is the same as for the Bell state and is equal to $\eta_H^{crit} = 2/(1 + \sqrt{2}) \approx 0.8284$~\cite{eberhard}.  Summing up, in this scenario, we need two detectors of efficiency $\eta_H > 0.8284$ and $N-2$ detectors of any efficiency greater than 0.

Note that in the case of using the four-qubit cluster state~$\rho_{Cl}$, one
detector (say the first one) can be completely blind ($\eta_1=0$).
It corresponds to the situation when we lose the first particle.
The resulting state is given by: $\rho_{\eta_1=0} = {\rm Tr}_1
\rho_{Cl} = 1/2(\proj{\psi_1} + \proj{\psi_2})$, where
$\ket{\psi_1} =\ket{1} (\ket{11} -\ket{00})/\sqrt{2}$ and
$\ket{\psi_2} = \ket{0}(\ket{00}+\ket{11})/\sqrt{2}$. When the
second observer chooses his/her projective measurement as $\Pi^+_2 =
\proj{0}$ (or $\Pi^+_2=\proj{1}$) the final two-qubit state is again the Bell state. So,
the critical efficiency $\eta_H^{crit}$ stays unchanged and is
equal to 0.8284.

We can even reduce the critical efficiency $\eta_H^{crit}$ in all presented examples to $2/3$ using the Eberhard inequality~\cite{eberhard}. In order to prepare the optimal state, the first projection $\Pi_1$ should be onto the asymmetric state $|\psi(\alpha)\rangle = \cos \alpha |0\rangle + \sin \alpha |1\rangle$. Then the resulting state is $\cos \alpha |00\rangle_{N-1,N} + \sin \alpha |11\rangle_{N-1,N}$ and for $\alpha \to 0$ we recover the optimal Eberhard result.

\subsection{Comparison with other methods}

\label{examp}

One can compare our results with those obtained by means of
another inequality. As an example let us consider the four-qubit
GHZ state and the corresponding WWW{\.Z}B inequalities \cite{ZB}
(or the cluster state and inequalities proposed by T\'oth et
al.~\cite{T}). When all detectors have the same efficiency $\eta$,
its critical value $\eta^{crit}$ is equal to $0.7706$ ($0.8205$),
so slightly less than in our case. However, if we set the
efficiency of the last two detectors to $0.8284$, then the
detection efficiency of the first two detectors have to be at
least $0.7169$ for the GHZ state and $0.8097$ for the cluster
state. These values are quite demanding compared to our method,
where $\eta_1 ,\eta_2 \geq 0$ for the GHZ state and  $\eta_1 > 0
,\eta_2 \geq 0$ for the cluster state.

\section{Duration of the experiment}

\label{time}

In order to reduce efficiency of the detectors one has to increase
the running time of the experiment to have the same statistics as
for the standard test. Obtaining experimentally the quantum value of the inequality we may take into account only such events when all detectors click and on the first $N-k$ particles ''+''-detector is the one which clicks. It happens with
probability of success $p_{succ} = p_1 \cdots p_{N-k}~
\eta_L^{N-k} \eta_H^{k}$. In the case of the standard scenario the
success probability is just $p_{succ}^{st} = \eta^N$. Such an experiment can be described by the Bernoulli distribution $B(r) = \binom{m}{r} p_{succ}^r (1-p_{succ})^{m-r}$, which tells us how many successes $r$ one can statistically obtain after $m$ trials.

\begin{figure}
    \centering
    \includegraphics[width=0.40\textwidth]{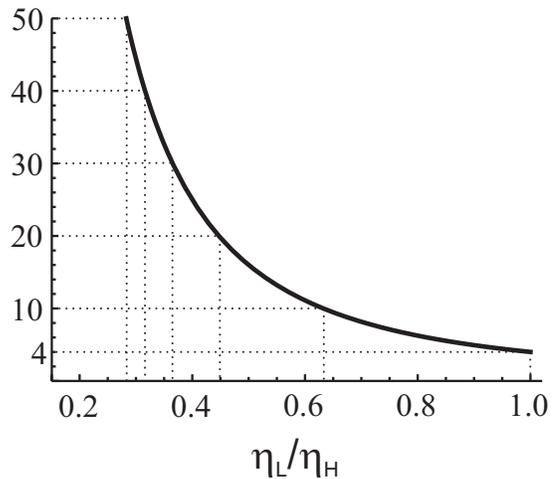}
    \caption{\label{fig-time} The trade-off relation between the ratio of low (projections) and high (standard inequality) value of detection efficiency and the growth of number of experimental trials $n'$. The $y$-axis is labeled by the ratio $n'$.}
\end{figure}

The average number of experimental trials to obtain $r$ successes with
the probability of a single success $p$ (the inverse problem) is given by $n_{avg} =
r/p$, which corresponds to the expectation value of the Pascal distribution. Assuming the same statistics $r$ for our scenario and the standard one, the ratio between the corresponding number of trials is:
\begin{equation}
n'=\frac{n_{avg}}{n_{avg}^{st}} =
\frac{p_{succ}^{st}}{p_{succ}} = \frac{\eta^N}{p_1 \cdots
p_{N-k} ~\eta_L^{N-k}  \eta_H^k}. \label{time-1}
\end{equation}
For simplicity let us consider the symmetric case, $\eta = \eta_H$. Then Eq. (\ref{time-1}) simplifies to
\begin{equation}
n' =    \frac{1}{p_1 \cdots p_{N_k}} \left( \frac{\eta_L}{\eta_H}\right)^{-N+k}.
\end{equation}
For further analysis we consider the example of the four-qubit GHZ
state from Sec. \ref{examp}, where we have the parameters $N=4$, $k=2$, and $p_1 = p_2 = 1/2$. In that case $n' = 4 (\eta_L / \eta_H)^{-1/2}$ (see also Fig.~\ref{fig-time}). It indicates for example that if the ratio $\frac{\eta_L}{\eta_H}$ reaches the value of $0.45$, one must perform $20$ times more experimental trials compared to the standard two-setting Bell test.

\section{Persistency of entanglement and nonlocality}
\label{pers}

We can also discuss our results  in the context of the {\em
persistency of entanglement} $P_e$ or the {\em presistency of nonlocality} $P_n$
\cite{cluster,P2}. In simple words, the persistency of entanglement
(nonlocality) of a given $N$-party state is the minimum number of
local measurements such that, for all measurement outcomes the
state is completely disentangled (local).

In our case, the situation is the opposite. Our goal is to find
the maximal number of local measurements (which ideally equals
to the number of ineffective detectors $N-k$) that a state remains
entangled (or nonlocal). In fact, in our case of entangled pure
$N$-party states, one can always find $N-2$ parties
such that the overall multipartite state violates a Bell inequality. This
argument is based on the findings in Refs.~\cite{PR,GG}. Namely,
it has been proven that considering an entangled pure $N$-party
state (of any dimensionality), it is always possible to carry out
specific projections on $N-2$ chosen particles such that the remaining
two particles end up in an entangled pure state for some special
outcomes of the $N-2$ projections. Since the projected
two-particle entangled state is pure, it can violate the CHSH
inequality due to the theorem of Gisin~\cite{gisin}.
Putting these together, we find that $N$-party pure entanglement
implies Bell nonlocality in the case of $N-2$ ineffective detectors. On the
other hand, Eberhard~\cite{eberhard} proves that any pure
entangled two-party state can violate the CHSH inequality with a
detection efficiency of at least 0.8284. This entails that it is enough to have two efficient detectors (with an efficiency of no less than 0.8284) in addition to the $N-2$ ineffective detectors in order to obtain Bell violation.


\subsection{The problem of damaged detectors}

One can also take a step forward and ask what is the maximal number of completely inefficient detectors that can be still tolerated in a Bell test to reveal nonlocality of a given state. The problem is related to the situation, when during the experiment some (unknown) detectors lost their function. In order to solve the problem we assume that the state is permutationally invariant and the potentially damaged detectors belong to the group of inefficient detectors ($i=1,\ldots, N-k$).

The quantum value of $I_{N-k+1,\dots,N}^{\eta_H}$ should be now calculated for the resulting state $\rho'_l$ after losing $l$ particles ($l < N-k$) and the projections of the first $N-k-l$ qubits:
\begin{equation}
\rho'_l = \frac{\Pi_{l+1}^+ \cdots \Pi_{N-k}^+~ {\rm Tr}_{1 \dots l} \rho_N ~\Pi_{l+1}^+ \cdots \Pi_{N-k}^+}{{\rm Tr}(\Pi_{l+1}^+ \cdots \Pi_{N-k}^+ ~ {\rm Tr}_{1 \dots l} \rho_N ~ \Pi_{l+1}^+ \cdots \Pi_{N-k}^+)}.
\end{equation}
Due to the above assumptions, it does not matter on which qubits we make the projection and on which the partial trace.

As in the previous examples, let $I_{N-1,N}$ be the CHSH inequality. The inequality is violated as long as we observe a non-zero overlap $\rho'_l$ with $|\psi^+\rangle = (|01\rangle + |10 \rangle)/\sqrt{2}$ state. Let us consider, as an example, the symmetric $N$-qubit Dicke state with $e$ excitations. After losing $l<N-2$ particles the state has the form~\cite{stockton}:
\begin{equation}
{\rm Tr}_{1 \dots l} \rho_N ={N \choose e }^{-1} \sum_{j=0}^l {l \choose j} {N-l\choose e-j} \rho_{D_{N-l}^{e-j}},
\end{equation}
where $\rho_{D_{N-l}^{e-j}} = |D_{N-l}^{e-j}\rangle \langle D_{N-l}^{e-j}|$.
After projecting ${\rm Tr}_{1 \dots l} \rho_N $ onto the state $|u \rangle_{1,\ldots,N-l-2} = |1\rangle_{l+1} \cdots |1\rangle_{l+u} |0\rangle_{l+u+1} \cdots |0\rangle_{N-2}$ we end up with the state $|\psi^+\rangle$
with probability
\begin{equation}
p_{|\psi^+\rangle } = 2 {l \choose e - u} {N \choose e }^{-1}.
\end{equation}
For all $N$ and $e$ one can find such $u$ that the probability of the $|\psi^+ \rangle$ fraction is greater then 0. We observe violation of the CHSH inequality even if some unknown detectors lost their function.


\section{Mixed states}
 \label{s-mixed}

In the previous sections we considered only pure states. It was
justified, because our main task was to find the critical value of
detection efficiency, which is optimal in the case of pure states.
However, it is also worth considering the more general situation of mixed
states, since experiments inevitably involve noise.
Assume that $\left|\psi_N \right\rangle$ is an $N$-particle pure
state and let us consider it with some white noise admixture
\begin{equation}
\rho = v\left|\psi_N \right\rangle \left\langle \psi_N \right| + \frac{1-v}{2^N} \openone,
\end{equation}
where $v$ is a mixing parameter called visibility of the state. We
now investigate such a mixture in the case of our inequality
(\ref{equ}). Let us assume that $Q_{\psi}(\eta_H)$ is the quantum
value of some (chosen) inequality for pure state and
$Q_{\rho}(\eta_H)$ is the quantum value of the same inequality when we add
white noise to the pure state. Then equation
\begin{equation}
v = \left(1 - 2^{N-k} \eta_L^{N-k} p_1 p_2 ... p_{N-k} \frac{Q_{\psi}(\eta_H)-L}{Q_{\rho}(\eta_H)-L} \right)^{-1} \label{v}
\end{equation}
provides a link between detection efficiency and the visibility of
our given state. Using this formula and taking $\eta_H = 1$, one
can find the critical visibility needed to violate our inequality.
The same formula also allows us to find the critical detection
efficiency in the noiseless case -- one just needs to assume $v=1$ and solve the
equation in that case.

\section{Conclusions}

We discussed the problem of closing the detection loophole in multipartite Bell tests
if only a limited number of highly efficient detectors are in our disposal, and the rest of the detectors can have arbitrary low efficiency. Our Bell tests are based on a family of $N$-partite Bell inequalities, where $k$ parties apply a $k$-party Bell inequality, whereas $N-k$ parties perform single projections on the remaining particles. Our construction can be applied in the case of several famous multipartite states, such as cluster states, GHZ states, and Dicke states, which have been realized in several photonic experiments. In particular, entangled polarization state of up to ten photons have been reported recently~\cite{ten}. We related the number of parties $k$ with efficient detectors to the notion of persistency of nonlocality, and also analyzed the trade-off relation between the required detection efficiencies and the time of Bell experiment. Finally, we analyzed the effect of noise in mixed states arising in Bell experiments. We believe that our inequalities, which are suited to a limited number of efficient detectors, may find useful information applications based on multipartite Bell nonlocality.

\section*{Acknowledgements}

K.K. is supported by NCN Grant No. 2014/14/M/ST2/00818. 
W.L. is supported by NCN Grant No. 2016/23/G/ST2/04273.
T.V. is supported by the National Research, Development and Innovation Office NKFIH (Grant Nos.  K111734, and KH125096).

\end{document}